# Domain Morphology, Electrocaloric Response, and Negative Capacitance States of Ferroelectric Nanowires Array

Anna N. Morozovska[1*], Oleksii V. Bereznykov[1], Maksym V. Strikha[2,3†], Oleksandr S. Pylypchuk[1], Zdravko Kutnjak[4], Eugene A. Eliseev[5‡], and Dean R. Evans[6§]

[1]Institute of Physics, National Academy of Sciences of Ukraine,
Nauky Avenue 46, 03028 Kyiv, Ukraine

[2]Taras Shevchenko National University of Kyiv, Faculty of Radiophysics, Electronics and Computer Systems, Pr. Akademika Hlushkova 4g, 03022 Kyiv, Ukraine,

[3]V. Lashkariov Institute of Semiconductor Physics, National Academy of Sciences of Ukraine, Nauky Avenue 41, 03028 Kyiv, Ukraine

[4] Slovenia Jožef Stefan Institute, Ljubljana, Slovenia

[5]Frantsevich Institute for Problems in Materials Science, National Academy of Sciences of Ukraine, Omeliana Pritsaka Street 3, 03142 Kyiv, Ukraine

[6]Zone 5 Technologies, Special Projects Division, San Luis Obispo CA 93401, USA

## Abstract

We analyzed the domain morphology, electrocaloric response, and negative capacitance states in a one-dimensional array of uniformly oriented, radial symmetric ferroelectric nanowires, whose spontaneous polarization is normal to their symmetry axis. The wires are densely packed between flat electrodes. Using finite element modeling based on the Landau-Ginzburg-Devonshire approach, electrostatics, and elasticity theory, we calculated the distributions of spontaneous polarization, domain structures, electric potential, electric field, dielectric permittivity, and electrocaloric response in the nanowires. Due to size and depolarization effects, the paraelectric and ferroelectric (poly-domain or single-domain) states of the wires can be stable, depending on their radius and the dielectric permittivity of the surrounding medium. It is demonstrated that dipole-dipole interaction between the nanowires determines the stability of the polar (or anti-polar) state in the array when the wire radius is significantly smaller than the critical size of the paraelectric transition in an isolated wire. We reveal that a large region of a mixed state, characterized by poly-domain ferroelectric states with nonzero average polarization inside each wire and zero average polarization of the whole array, can be stable. The mixed state is nearly absent in an isolated wire or at a significant distance between the wires, being inherent to their dense packing only. By selecting the dielectric permittivity of the surrounding medium and the nanowire radius, one can maximize the negative capacitance effect in the capacitor with densely packed wires. It is also possible to achieve maximal enhancement of the electrocaloric response due to size effects in the wires. The underlying physics of the predicted enhancement is the combined action of size effects and the long-range electrostatic interactions between the ferroelectric dipoles in the nanowires and the image charges in the electrodes.

---

[*] corresponding author, e-mail: anna.n.morozovska@gmail.com
[†] corresponding author, e-mail: maksym.strikha@gmail.com
[‡] corresponding author, e-mail: eugene.a.eliseev@gmail.com
[§] corresponding author, e-mail: dean.evans92@gmail.com

## I. INTRODUCTION

Colloids and composite films and containing ferroelectric (FE) nanoparticles of various shapes and sizes are unique model systems for fundamental studies of interfacial and size effects [1, 2, 3], as well as dipole-dipole interactions in ensembles of nanoparticles [4, 5]. These nanomaterials attract practical interest because of their remarkable promise for energy storage [6, 7, 8], piezoelectric [9], pyroelectric [10, 11], and electrocaloric (EC) applications [12, 13].

It was shown theoretically that finite size effect can be the main reason for the strong increase of pyroelectric and electrocaloric responses in single-domain FE nanorods [14, 15] and nanospheres [13, 16, 17]. It has also been demonstrated that the negative capacitance (NC) state, introduced and observed in layered FE films [18, 19], can emerge in films with semiconductor [20] or FE nanoparticles [5].

In this work, we consider a one-dimensional array of uniformly oriented radially symmetric FE nanowires densely packed between flat electrodes and possessing spontaneous polarization normal to their axis of symmetry. This geometry, unlike that considered in our previous theoretical works [14, 15], represents a practically important case realized in experiments [21, 22, 23] and enabled by existing preparation methods [24, 25]. We analyze how size and screening effects, the surrounding medium's dielectric permittivity, and long-range dipole-dipole interactions between neighboring nanowires affect polarization, domain structure morphology, dielectric permittivity, and electrocaloric response. We identify the conditions that enable the NC state in the wires and maximize their EC response, placing special emphasis on the underlying physics of the predicted effects.

## II. PROBLEM FORMULATION AND TYPICAL CASES

Let us consider the case of a linear array of uniformly oriented FE nanowires with a circular cross-section of radius $R$. The wires are densely packed: the distance between their sidewalls is less than a lattice constant (set to 2 Å in the calculations). The spontaneous polarization $P_s$ of each wire is directed along the $Z$ axis normal to the wire axis $Y$ (see designations in **Fig. 1(a)**). With an ultra-thin physical gap (set to 1 Å in the calculations), each wire nearly contacts the top and the bottom conducting electrodes. The space between the wires is filled by an elastically soft dielectric medium with the relative dielectric permittivity $\varepsilon_S$. Free charges are absent in both the wires and the dielectric material. This geometry maximizes the dipole-dipole interaction between FE dipoles of neighboring nanowires. It is also of practical importance [21, 22], as it enables straightforward evaluation of the system capacitance (i.e., capacitance between the electrodes) and the EC response.

A periodic voltage $U(t) = U_0 \sin(\omega t)$ is applied between the electrodes. The electric potential $\varphi$ is set to $U(t)$ at the top electrode, while the bottom electrode is grounded ($\varphi = 0$). The applied voltage is time dependent; a slowly varying AC bias is used to study the domain formation and the EC response. The bias frequency $\omega$ is assumed to be much smaller than the inverse Landau-Khalatnikov relaxation time of polarization, $\tau_{LK}$, in order to simulate the so-called quasi-static hysteresis loops. Specifically, the frequency is set to $\omega = 10^{-4}(2\pi/\tau_{LK})$.

Distributions of the electric potential and field, polarization, emerging domain structures, dielectric permittivity, and EC response of the nanowires are computed by finite element modeling (FEM) using the phenomenological Landau-Ginzburg-Devonshire (LGD) approach, together with electrostatic equations and elasticity theory. Details of the calculations, the system of nonlinear LGD equations with boundary conditions, and material parameters of BaTiO$_3$ are given in **Appendix A** [26]. We performed FEM for a range of wire radii ($R = 2 - 20$ nm) and surrounding medium permittivities ($\varepsilon_S = 1 - 300$). The wires are assumed to be sufficiently long such that their polarization and all physical quantities are independent of the coordinate Y.

In what follows, we focus on two limiting cases: (1) when the permittivity $\varepsilon_S \leq 10$ is much smaller than the "reduced" permittivity $\varepsilon_f$ of a bulk FE material, and (2) when $\varepsilon_S$ is much larger than $\varepsilon_f$. The reduced permittivity is defined as $\varepsilon_f = \sqrt{\varepsilon_{11}^f \varepsilon_{33}^f}$, which is about 200 for tetragonal BaTiO$_3$ at room temperature. The condition $\varepsilon_S \ll \varepsilon_f$ corresponds to a surrounding liquid or polymer (such as liquid crystal, heptane, or polymers like PVB, PVFD, and P(VDF-TrFE)), whereas the condition $\varepsilon_S > \varepsilon_f$ corresponds to a paraelectric (PE) material (e.g., for SrTiO$_3$) or a relaxor-type dielectric. These two cases correspond to fundamental differences in the distribution of the external electric field between the electrodes. The external field is concentrated outside the wires when $\varepsilon_S \ll \varepsilon_f$, and inside them when $\varepsilon_S \gg \varepsilon_f$. In the former case, the nanowires may exhibit a PE state or domain splitting, while in the latter case they tend to remain single-domain. Size and depolarization effects determine whether the wires are in the PE, poly-domain (PD) FE, or single-domain (SD) FE state; the stability of these states depends on the wire radius $R$ and the surrounding dielectric permittivity $\varepsilon_S$. Both factors, $R$ and $\varepsilon_S$, determine the phase state of the wire because they are coupled through the depolarization field produced by the wires. The field of a solitary single-domain wire can be roughly estimated as $E_3^d = -\frac{1}{\varepsilon_b + \varepsilon_S} \frac{P_3}{\varepsilon_0}$, where $\varepsilon_0$ is the vacuum permittivity, $\varepsilon_b$ is the relative background permittivity of the FE [27], and the wire polarization $P_3$ is $R$-dependent. Note that the proximity of the wires to the conducting electrodes (to the greatest extent) and electrostatic cross-interaction between the wires (to a lesser extent) change the depolarization field induced by the polarization of the wires.

**Figure 1(a)** shows typical equilibrium polarization states in FE nanowires for poly-domain ($\varepsilon_S \ll \varepsilon_f$) and alternating single-domain ($\varepsilon_S > \varepsilon_f$) polarizations. In the latter case, each wire is single-domain, and neighboring wires have opposite polarization directions (like an array of dipoles stacked head-to-tail). FE domains appear in the nanowires at small values of $\varepsilon_S$ when their radius $R$ exceeds the critical value $R_{cr}$ [28, 29]. Under these conditions, the domain formation minimizes the energy of the depolarization field. The wire is in the PE phase for $R < R_{cr}$ (not shown in the figure). The "mixed" domain states, characterized by nonzero average polarization inside each wire but zero average polarization of the whole array ($10 \ll \varepsilon_S < \varepsilon_f$), are shown in the middle image in **Fig. 1(a)**. The mixed state is denoted "SD+PD FE", because it can be imagined as a mixture of the single-domain and poly-domain states coexisting inside a single wire. Note that the range $10 \ll \varepsilon_S < \varepsilon_f$ corresponds to solid dielectrics, such as rutile with $\varepsilon_S \cong 90 - 120$ and to polar liquids with $\varepsilon_S \cong 80$.

Large values of $\varepsilon_S$ (e.g., close to or larger than $\varepsilon_f$) support the stability of the single-domain state in nanowires of small radius. However, in this case, the neighboring wires are counter-polarized to minimize the electrostatic energy of the whole system (shown in the bottom image in **Fig. 1(a)**). A large $\varepsilon_S$ also implies a smaller dielectric mismatch between the wire and its surroundings, which makes it easier for bound charges to accumulate at the poles near the electrodes. As shown below, this effect enhances the spontaneous polarization near the wire poles.

Typical equilibrium polarization distributions (cross-section view) calculated in BaTiO$_3$ nanowires for different $R$ and $\varepsilon_S$ are shown in **Fig. 1(b)**. The structures correspond to the "off-field" relaxation of the randomly small initial distribution of polarization inside the wires at $U = 0$. Hereinafter, we refer to the equilibrium configuration as the ground state "0". The polarization dynamics are then calculated under a periodic bias voltage, $U$, and the resulting "on-field" relaxation is analyzed. The four upper images in **Fig. 1(b)** correspond to a fixed permittivity ($\varepsilon_S = 10$) and decreasing radius $R$ (from 20 nm to 5 nm), illustrating a gradual transition from the PD FE to the PE state as $R$ decreases. The three lower images in **Fig. 1(b)** correspond to a fixed radius ($R = 5$ nm) and increasing $\varepsilon_S$ (from 10 to 300). They illustrate the transitions from the PE state to the SD+PD FE state and finally to the SD FE state.

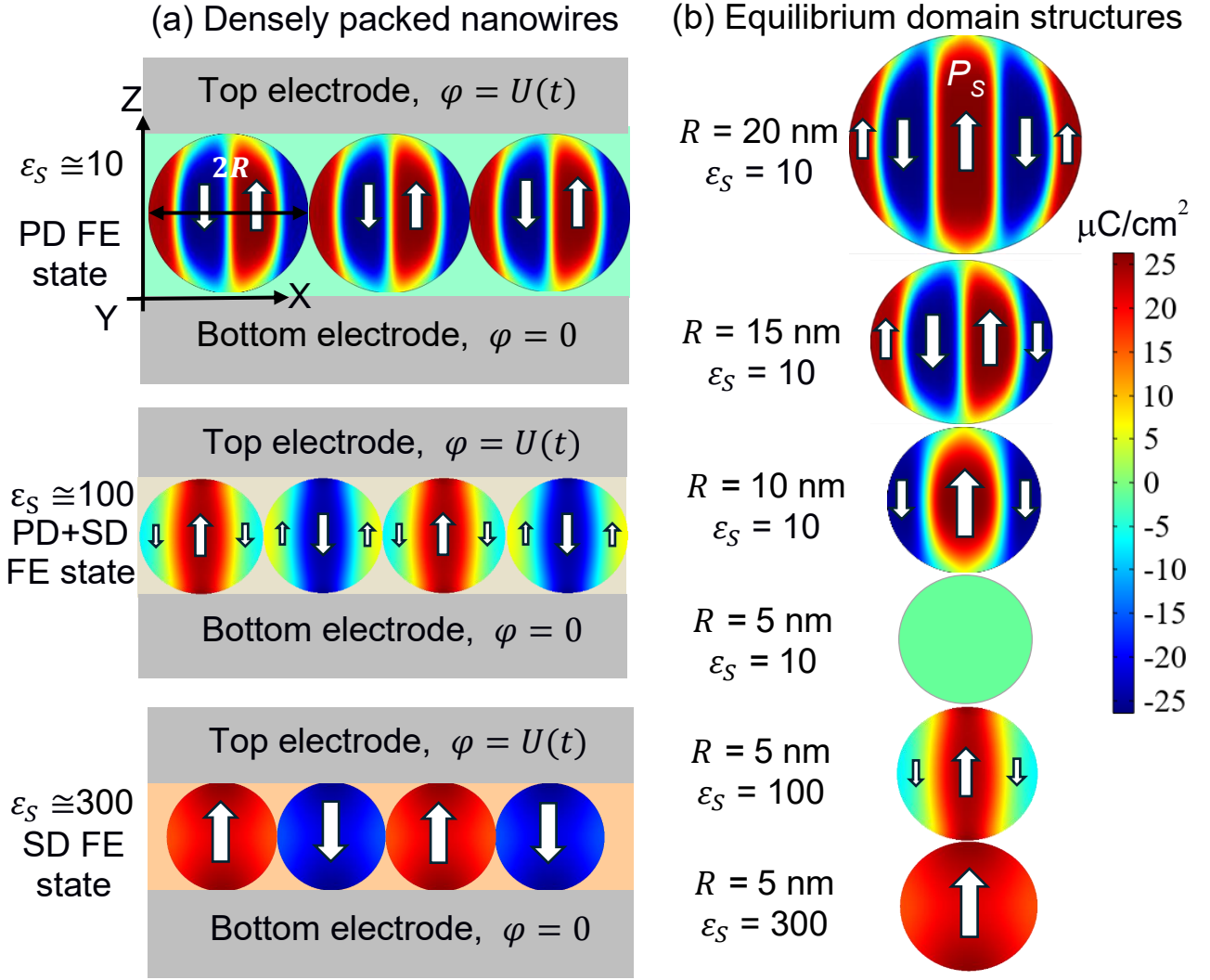

**FIGURE 1. (a)** XZ-section view of a one-dimensional array of uniformly oriented FE nanowires with a circular cross-section of radius $R$. The Y axis of each nanowire is normal to the plane of the figure. The spontaneous polarization of the wires is directed along the axis Z. The array dimension is measured along the X axis. The wires are densely packed: their sidewalls nearly touch, separated by an ultra-thin gap of less than a lattice constant, and each wire is in contact with the top and bottom electrodes. The space between the wires is filled with a dielectric medium with relative dielectric permittivity $\varepsilon_S$. Typical cases of the equilibrium poly-domain (the top image for $\varepsilon_S \cong 10$), mixed single-domain and poly-domain (the middle image for $\varepsilon_S \cong 100$), and alternating single-domain (the bottom image for $\varepsilon_S \cong 300$) states of the nanowires are shown. **(b)** Equilibrium polarization distributions (XZ cross-section view) calculated in BaTiO$_3$ nanowires for different $R$ and $\varepsilon_S$; values are indicated next to the images. White arrows show the polarization direction along the Z axis in the nanowire. The images correspond to the "off-field" relaxation of the polarization at $U = 0$ and $T = 298$ K.

### III. RESULTS AND DISCUSSION

The phase state of densely packed BaTiO$_3$ nanowires was analyzed as a function of the wire radius $R$ and the relative dielectric permittivity of the environment $\varepsilon_S$, for zero bias ($U = 0$) at room temperature $T = 298$ K. The schematic diagram in **Fig. 2** illustrates the different phase states (PE,

PD FE, and SD FE) that emerge under varying conditions. The phase state of the nanowires is determined by the minimum of the LGD free energy. For $U = 0$, the main contributions to the LGD free energy are the positive depolarization field energy, the polarization gradient energy (including the domain wall energy), and the Landau energy, which is negative in FE states and zero or positive in the PE phase. Similarly to the well-studied case of an isolated nanowire, the PE state is stable for radii smaller than the critical radius $R_{cr}$, where $R_{cr}$ decreases with increasing $\varepsilon_S$. The decrease of $R_{cr}$ is related to an increase in the depolarization field energy, which is proportional to $\frac{1}{\varepsilon_S}$. At $\varepsilon_S > \varepsilon_f$, the value of $R_{cr}$ becomes smaller than 2 nm, i.e., below the applicability limit of the continuous-media LGD approach.

The SD FE is stable for large $\varepsilon_S$ (> 150) and relatively small radii (< 5 - 7 nm). The stability of the SD FE state is conditioned by the tendency of the nanowires to minimize their free energy through a reduction of the positive domain wall energy. However, this condition holds only when the depolarization field contribution is relatively small, as domain splitting strongly reduces the depolarization field energy. Since the depolarization field energy is proportional to $\frac{1}{\varepsilon_S}$, the stability region of the SD FE state broadens with increasing $\varepsilon_S$. The stability of the SD FE at small sizes (~5 - 10 nm) is in line with earlier studies of size effects in BaTiO$_3$ nanoparticles [30, 31, 32]. As depolarization effects are strongly reduced with increasing $\varepsilon_S$, the critical radius $R_{cr}$ approaches the fundamental limit for the disappearance of ferroelectricity, which corresponds to the correlation volume (~ 10 lattice constants) that defines the lower limit for stable ferroelectric order [33].

The coexistence region of the SD FE and PD FE states (denoted as "SD+PD FE") is characterized by a nonzero average polarization of each nanowire and lies between two dashed curves in **Fig. 2**. This nonzero average indicates that the total areas of up-and down-polarized domains in the XZ cross-section of the wire are unequal. At the same time, the overall average polarization of the entire array remains zero, as illustrated in the middle image in **Fig. 1(a)**. Note that the alternating pattern of nanowire polarizations, stabilized by dipole–dipole interactions, is perfectly balanced due to the long-range nature of the dipole forces. The SD+PD FE state gradually transforms to the PD FE state with a decrease in $\varepsilon_S$ and/or an increase in $R$; it gradually transforms to the SD FE state with an increase in $\varepsilon_S$ and/or decrease in $R$.

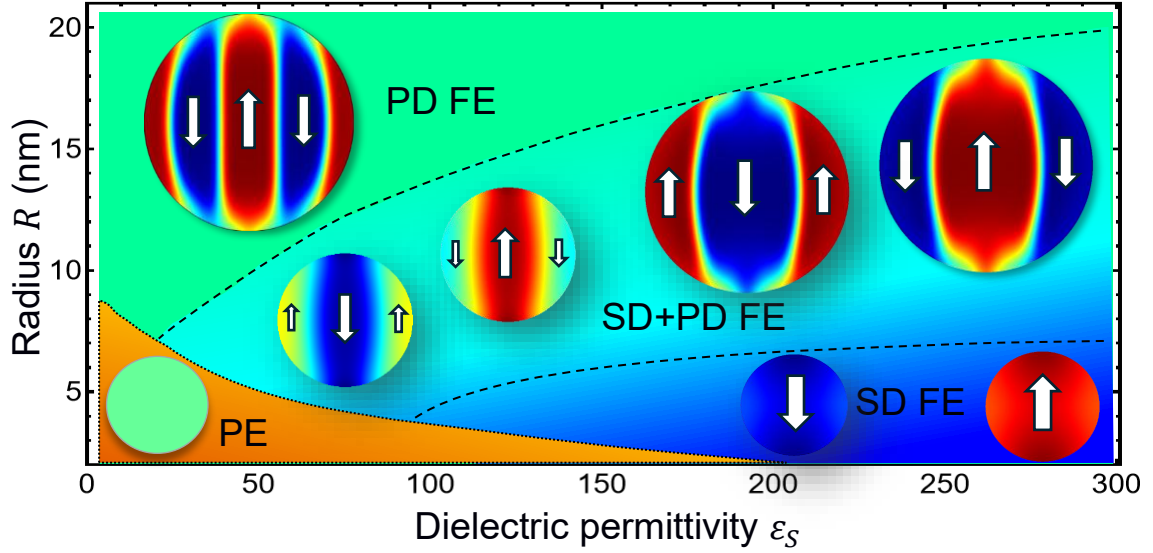

**FIGURE 2.** The ground state of BaTiO₃ nanowires as a function of wire radius $R$ and relative dielectric permittivity of the environment $\varepsilon_S$ for $U = 0$ and room temperature $T = 298$ K. The abbreviations PE, PD FE, and SD FE denote the paraelectric, polydomain ferroelectric, and single-domain ferroelectric states, respectively. The abbreviation SD+PD FE denotes the coexistence region of the SD FE and PD FE mixed states. Color images with white arrows schematically illustrate the spontaneous polarization distribution and direction in the XZ cross-section of the nanowire for paraelectric, poly-domain, and single-domain (up- and down-polarized) states. The color scale for polarization is the same as in **Fig. 1(b)**: positive values are red, negative values are blue, and zero values are light green.

The average spontaneous polarization ($P_S^{\text{avg}}$) of an individual BaTiO₃ nanowire, calculated over its XZ cross-section, is presented in **Fig. 3(a)** as a function of wire radius $R$ and dielectric permittivity of the environment $\varepsilon_S$. In the case of counter-polarized single-domain nanowires (as shown at the bottom of **Fig. 1(b)**), an up-polarized wire is defined as one with a positive average polarization. We would like to emphasize that the average polarization of the entire array is zero in the ground state, corresponding to the system's minimal electrostatic energy. The dark-violet region with $P_S^{\text{avg}} = 0$ corresponds to the PE state. The dark-blue region, where $P_S^{\text{avg}}$ is small, corresponds to the PD FE state. The large region, with the color gradually changing from blue to orange, corresponds to the SD+PD FE state, where $P_S^{\text{avg}}$ increases with an increase in $R$ and/or $\varepsilon_S$. The SD+PD FE state transforms gradually to the SD FE state with a decrease in $R$ and an increase in $\varepsilon_S$. The color of the SD FE state changes from yellow to red with an increase in $\varepsilon_S$ from 100 to 300, which corresponds to an increase in $P_S^{\text{avg}}$ from 15 μC/cm² to 22 μC/cm².

The maximal spontaneous polarization ($P_S^{\text{max}}$) of the BaTiO₃ nanowire as a function of $R$ and $\varepsilon_S$ is shown in **Fig. 3(b)**. The $P_S^{\text{max}}$ refers to the local maximum of polarization within a given nanowire cross-section, calculated for fixed $R$ and $\varepsilon_S$. The dark-violet region with $P_S^{\text{max}} = 0$ corresponds to the PE state. The large region, with color continuously changing from orange to red, contains PD FE,

SD+PD FE, and SD FE states, where $P_S^{\text{max}}$ increases gradually with an increase in $\varepsilon_S$. The dashed black curves separate these three states for the reader's convenience, since their boundaries are not distinguishable by the color coding in **Fig. 3(b)**.

The minimal spontaneous polarization ($P_S^{\text{min}}$) of the BaTiO$_3$ nanowire as a function of $R$ and $\varepsilon_S$ is shown in **Fig. 3(c)**. $P_S^{\text{min}}$ refers to the local minimum of polarization within a given nanowire cross-section calculated for fixed $R$ and $\varepsilon_S$. The olive-green region with $P_S^{\text{min}} = 0$ corresponds to the PE state. The trapezoidal-shaped violet-blue region, where $P_S^{\text{min}}$ is large and negative, namely $P_S^{\text{min}} = -P_S^{\text{max}}$, corresponds to the PD FE state. The stripe-shaped region, with the color gradually changing from dark-blue to olive-green, corresponds to the SD+PD FE state, where $P_S^{\text{min}}$ increases from negative values to zero as $\varepsilon_S$ increases and/or $R$ decreases. The wedge-shaped region, with color gradually changing from olive-green to red, corresponds to the SD FE state, where $P_S^{\text{min}}$ increases from zero to positive values as $\varepsilon_S$ increases and/or $R$ decreases. The boundary between the SD+PD FE and the SDFE states, corresponding to $P_S^{\text{min}} = 0$, is shown by the dashed black curve in **Fig. 3(c)**.

To highlight the role of collective effects, it is useful to compare phase diagrams of the nanowire array, shown in **Fig. 3(a-c)**, with that of an isolated nanowire embedded in a flat capacitor with the same $\varepsilon_S$ (see **Fig. S1** in Ref. [26]). The diagram of an isolated nanowire is well-known [28-29] and thus not shown in the main text. Due to the dipole-dipole interaction between the nanowires, the polar (or anti-polar) state of the array remains stable for wire radii $R$ that are significantly smaller than the critical size for the paraelectric transition in the isolated wire (compare **Figs. 2** and **S1**). Another key distinction is the emergence and stability of a large SD+PD FE region in the nanowire array which is nearly absent in the case of the isolated wire. The phase diagram of an isolated wire contains the PE, PD FE, and SD FE states, whose stability ranges are determined by $R$ and $\varepsilon_S$ values. The long-range dipole–dipole forces drive the array of nanowires toward zero net polarization, thereby stabilizing the mixed SD+PD FE state of the array, which minimizes the system's electrostatic energy.

To clarify the origin of these differences in polarization behavior between the array and the isolated nanowire, we next examine how local field effects — particularly surface curvature and image charges at the electrodes — modify the polarization distribution within individual wires. In particular, the spatial distribution of polarization may exhibit a local enhancement near the electrodes. Namely, **Fig. 3(b)** and **3(c)** show that the minimal and maximal polarizations exceed the bulk value of 26 μC/cm$^2$ [34] for $\varepsilon_S > \varepsilon_f$, while the average polarization is less than the bulk value for BaTiO$_3$ (see **Fig. 3(a)**). A detailed investigation showed that the polarization enhancement is localized near the wire "poles", specifically near the "electrode-facing" poles, where the nanowire's surface curvature intensifies the local depolarization field and cooperates with the image charges at the electrodes. This phenomenon is observed only for sufficiently large values of the permittivity $\varepsilon_S \gtrsim \varepsilon_f$ and for wire regions located very close to the electrodes. As $\varepsilon_S$ decreases or the gap between the wires and the electrodes increases, the polarization enhancement disappears. Large $\varepsilon_S$ reduces the dielectric mismatch and strongly supports the stability of the SDFE state, making it more favorable for bound charges to accumulate at the electrode-facing poles, thereby enhancing local polarization. However, very large or "colossal" values of $\varepsilon_S$ (e.g., $\varepsilon_S \gg 10^3$) almost suppress the depolarization field in the nanowire and make its polarization distribution homogeneous (approaching the bulk value), resulting in the vanishing of polarization enhancement near the electrode-facing poles. Thus, an optimal range of $\varepsilon_S$ for the observation of polarization enhancement should exist. According to FEM results, this range corresponds to $\varepsilon_f < \varepsilon_S < \varepsilon_f \cdot \varepsilon_b$.

The enhancement is accompanied by the highly non-uniform distribution of the internal electric field (or depolarization field) inside and outside the single-domain nanowires. In contrast to single-domain nanowires, planar layered single-domain systems (such as thin-film stacks) exhibit a uniform depolarization field within each individual layer. As a result, it does not enhance the polarization; on the contrary, it tends to equalize the polarization between layers [35]. In contrast to the collective dipole-dipole effects governing the array-scale phase stability, the local polarization enhancement in single-domain nanowires is governed primarily by surface curvature and electrostatic interaction of the bound charges near the wire surface with image charges at the electrodes (i.e., image-charge interactions).

In contrast to an isolated nanowire, where polarization enhancement is primarily governed by surface curvature and image-charge effects, in densely packed arrays it is dominated by long-range dipole–dipole interactions between neighboring wires. These collective effects, together with the size and environmental permittivity, determine the overall dependences of average spontaneous polarization $P_S^{\mathrm{avg}}$, linear effective dielectric permittivity $\varepsilon_{eff}$, and EC coefficient $\Sigma$ on the nanowire radius $R$, as shown in **Fig. 3(d), 3(e),** and **3(f),** respectively. The values of $\varepsilon_S$ are 10, 30, and 300 for the black, red, and blue curves, respectively. These values correspond to high-k polymers such as PVDF and P(VDF-TrFE) with $\varepsilon_S \cong 10 - 30$, and paraelectric SrTiO$_3$, which has a relative dielectric permittivity of about 300 at room temperature. As seen in **Fig. 3(d-f),** the values of $P_S^{\mathrm{avg}}$, $\varepsilon_{eff}$, and $\Sigma$ increase with increasing $\varepsilon_S$ (compare the black, red, and blue curves). Note that the state of the nanowire array changes with an increase in $R$ at a fixed $\varepsilon_S$, and the corresponding transitions occur at the interaction points of white dotted vertical lines ($\varepsilon_S = 10, 30,$ and 300) with the black dashed phase-boundary curves in **Fig. 3(a).** For $\varepsilon_S = 10$, the array state changes from PE to PD FE as $R$ increases from 2 to 10 nm. For $\varepsilon_S = 30$, the array state evolves from PE to SD+PD FE and then to PD FE. For $\varepsilon_S = 300$, it changes from the SD FE state to the SD+PD FE state and finally to the PD FE state as $R$ increases from 2 to 10 nm.

The black ($\varepsilon_S = 10$) and red ($\varepsilon_S = 30$) curves in **Fig. 3(d-f)** reveal a size-induced phase transition from the FE to the PE phase as the nanowire radius decreases below the critical value $R_{cr}$, which is approximately 8.5 nm and 6.5 nm, respectively. The spontaneous polarization and EC response vanish at $R \leq R_{cr}$; their step-like onset corresponds to $R = R_{cr}$. The effective permittivity $\varepsilon_{eff}$ exhibits a pronounced step-like maximum at $R = R_{cr}$, being consistent with a first-order phase transition for black and red curves. For $R > R_{cr}$, the spontaneous polarization increases slightly, while the EC coefficient decreases slightly for these curves. In general, the size-induced phase transition is observed only for $\varepsilon_S < \varepsilon_f$.

Due to the high degree of dielectric screening, the size-induced transition to the PE phase is absent for $\varepsilon_S > \varepsilon_f$, which is illustrated by the corresponding behavior of the blue curves in **Fig. 3(d-f)** calculated for $\varepsilon_S = 300$ and $\varepsilon_f = 200$. In contrast to the lower $\varepsilon_S$ cases, at $\varepsilon_S = 300$ the spontaneous polarization $P_S^{\mathrm{avg}}$ decreases slightly with increasing $R$.

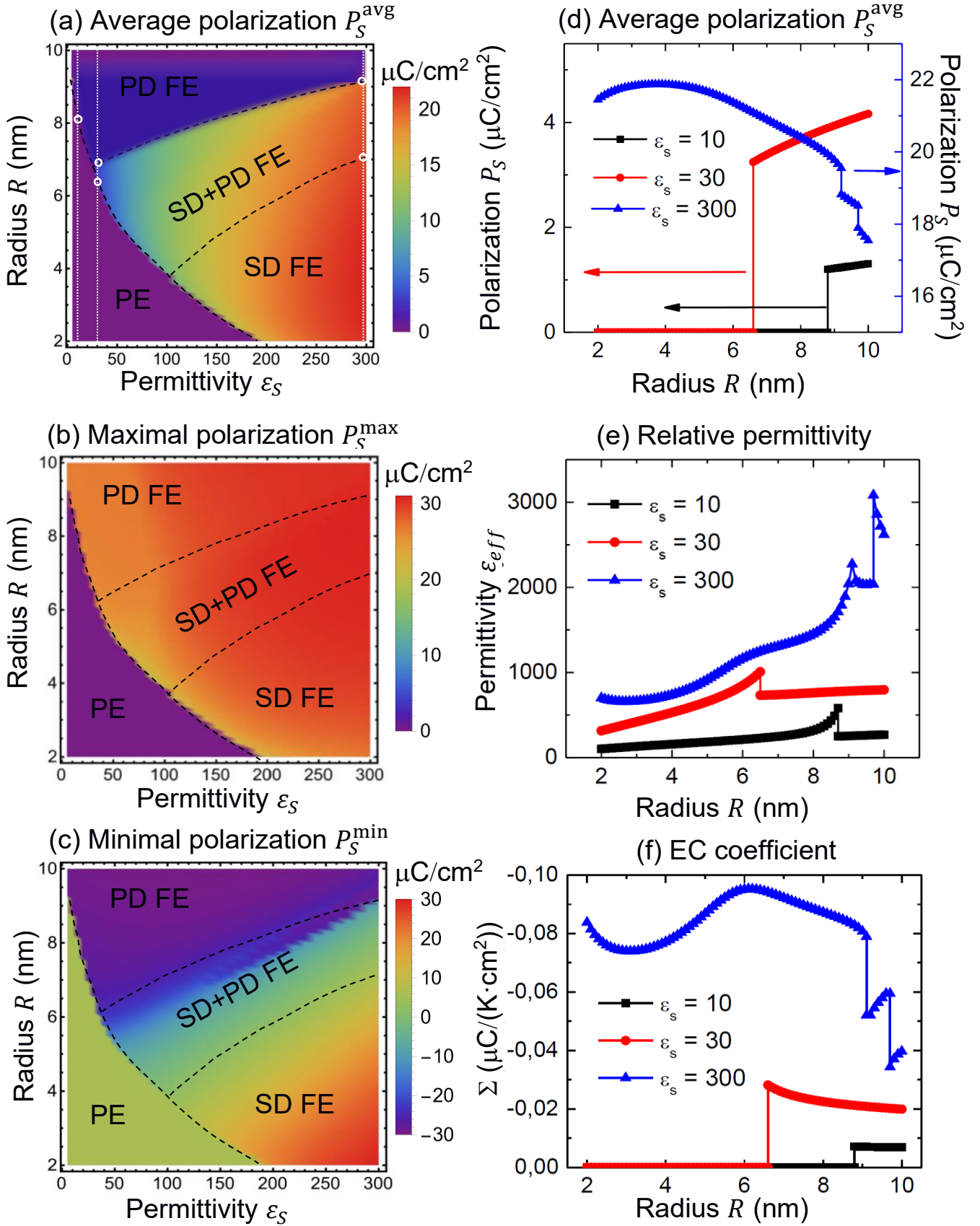

**FIGURE 3.** **(a)** Average spontaneous polarization ($P_S^{\mathrm{avg}}$), **(b)** maximal spontaneous polarization ($P_S^{\mathrm{max}}$), and **(c)** minimal spontaneous polarization ($P_S^{\mathrm{min}}$) of the individual BaTiO₃ nanowire as a function of wire radius $R$ and dielectric permittivity of environment $\varepsilon_S$. White dotted vertical lines in panel (a) correspond to $\varepsilon_S = 10$, 30, and 300. The color scales on the right show the values of $P_S^{\mathrm{avg}}$, $P_S^{\mathrm{max}}$, and $P_S^{\mathrm{min}}$ in μC/cm². Dependences of the **(d)** average spontaneous polarization $P_S^{\text{avg}}$, **(e)** linear dielectric permittivity $\varepsilon_{eff}$, and **(f)** EC coefficient $\Sigma$ **(f)** on the wire radius $R$ calculated for $\varepsilon_S = 10$ (black curves), 30 (red curves), and 300 (blue curves). Diagrams **(a-c)** correspond to the "off-field" relaxation of the polarization from small random fluctuations. Plots **(d-f)** were obtained after applying a quasi-static periodic voltage with an amplitude of 0.1 V, followed by its removal and subsequent polarization relaxation. The temperature $T = 298$ K.

We next explore the conditions that yield a maximal EC response by analyzing the hysteretic properties of polarization, dielectric, and EC responses for $R \approx 2$ nm and $R \approx 6$ nm at $\varepsilon_S = 300$, since these parameters correspond to the maximal values of $\Sigma$ in **Figs. 3(f)**. The corresponding hysteresis loops of the electrode charge density, for the case where the BaTiO$_3$ nanowires array is placed between the electrodes, are presented in **Figs. 4(a)** and **5(a)**. The charge loops exhibit a pronounced rectangular shape with bistable remanent polarization states "+1" and "-1", indicated on the loops. The coercive voltages are 20 mV (for $R = 2$ nm) and 48 mV (for $R = 6$ nm). The corresponding effective coercive fields are 500 kV/cm (for $R = 2$ nm) and 800 kV/cm (for $R = 6$ nm).

Hysteresis loops of the effective dielectric permittivity $\varepsilon_{eff}$, shown in **Figs. 4(b)** and **5(b),** have a very sharp maxima (>5000) at the coercive voltages. A notably feature is that the linear effective permittivity ($\varepsilon_{eff}$ at $U = 0$ and $\varepsilon_S = 300$) exceeds 600 for $R = 2$ nm and 1200 for $R = 6$ nm nanowires. Since $\varepsilon_{eff}$ is 3 - 4 times higher than the ambient permittivity $\varepsilon_S = 300$ and the bulk value $\varepsilon_f = 200$, the contribution of BaTiO$_3$ nanowires, whose volume fraction $\mu$ is about 79 %, results in a pronounced enhancement of the effective dielectric response. The enhancement does not follow the simple mixture law, $\varepsilon_{eff} = \mu\varepsilon_f + (1 - \mu)\varepsilon_S$; instead it is induced by size effects in the nanowires, where the local dielectric permittivity $\varepsilon_{loc}$ (i.e., the effective dielectric response within an individual nanowire) is much greater than the permittivity $\varepsilon_f$ of bulk BaTiO$_3$ at room temperature.

Hysteresis loops of the EC coefficient $\Sigma$, shown in **Figs. 4(c)** and **5(c)**, reveal a significant enhancement of the EC response near the coercive voltages, followed by a gradual decrease with a further increase of voltage. Near the coercive voltages, $\Sigma$ is greater than 0.3 μC/(Kcm$^2$) for $R = 2$ nm and greater than 0.2 μC/(Kcm$^2$) for $R = 6$ nm. Even more important is the enhancement of $\Sigma$ at small voltages. In the low voltage limit ($U \to 0$), $\Sigma$ reaches 0.08 μC/(Kcm$^2$) for $R = 2$ nm and 0.098 μC/(Kcm$^2$) for $R = 6$ nm. These values are 4-5 times greater than those for a bulk BaTiO$_3$, which is about 0.02 μC/(Kcm$^2$) at small voltages. The strong enhancement of the EC results from the combined influence nanowire size effect, electrostatic interactions with electrode image charges, and dipole-dipole interactions between neighboring wires.

The panels in **Figs. 4(d)** and **5(d)** (from the top to bottom) show the distributions of spontaneous electric polarization $P_S(\vec{r})$, electric potential $\varphi(\vec{r})$, z-component of the internal electric field $E_z(\vec{r})$, local dielectric permittivity $\varepsilon_{loc}(\vec{r})$, and EC response $\Sigma(\vec{r})$. These quantities are shown for BaTiO$_3$ nanowires in the "+1" (right column) and "-1" (left column) states, as indicated on the charge loop. The nanowires are single-domain for large $\varepsilon_s > \varepsilon_f$ and mostly poly-domain for small $\varepsilon_s < 50$ (see **Figs. S2** and **S3** in Ref. [26]).

The electric potential $\varphi$ reaches a maximum near the top pole and a minimum near the bottom pole of the nanowires, where they contact the electrodes. The maximal and minimal values of $\varphi$ increase with wire radius $R$: $\varphi$ is $\pm 30$ mV for $R = 2$ nm and $\pm 60$ mV for $R = 6$ nm. The potential decreases slightly within the ultra-thin gap between the wire pole and the electrode surface, an effect caused by the image charges induced in the electrode. The potential is zero in the axial symmetry plane of the nanowires.

At $U = 0$, the internal electric field corresponds to the depolarization field and is concentrated at the contact regions of the nanowires with the electrodes, where it reaches the maximal values due to the interaction of surface polarization (bound charges) with the image charges in the electrodes. Because of this, the behavior of $E_z(\vec{r})$ inside the wires is governed by the spontaneous polarization gradient: it reaches its maximum where the gradient is strongest and its minimum where the gradient is the weakest.

For $R = 2$ nm, the local dielectric permittivity $\varepsilon_{loc}(\vec{r})$ is positive inside the nanowire, minimal at the central, intermediate along the polar diameter in contact with the electrodes, and maximal near lateral contacts with neighboring wires. The spatial distribution of $\varepsilon_{loc}(\vec{r})$ inside the wires is governed by the distribution of spontaneous polarization: it is maximal where the spontaneous polarization varies most strongly and minimal where the polarization is nearly uniform. However, $\varepsilon_{loc}(\vec{r})$ is negative outside the wires near the contact line between neighboring wires. For $R = 6$ nm, $\varepsilon_{loc}(\vec{r})$ is also negative within a relatively large central region of the nanowires and in smaller regions outside the wire near the contact line between the neighboring wires.

Regions of negative permittivity are of particular interest, as they may correspond to the emergence of the NC state. **Figs. 4(e)** and **5(e)** highlight the spatial regions where $\varepsilon_{loc}(\vec{r})$ is negative. The appearance of the NC state, as well as the size, spatial location, and distribution of NC regions and the local permittivity within them, are determined by the wire radius $R$ (at fixed $\varepsilon_S$). For $R = 2$ nm, the NC state is confined to a thin equatorial region near the contact point of the neighboring nanowires. In contrast, for $R = 6$ nm, the NC state extends into a relatively large area in the central part of the nanowires. According to thermodynamic theory, the total capacitance of the system is positive in thermal equilibrium. This requirement is consistent with the conventional shape of the charge density hysteresis loops, which show no regions with a negative slope $dQ/dU < 0$ (as shown in **Figs. 4(a)** and **5(a)**), and with the positive values of the effective dielectric permittivity $\varepsilon_{eff}$ (as shown in **Figs. 4(b)** and **5(b)**).

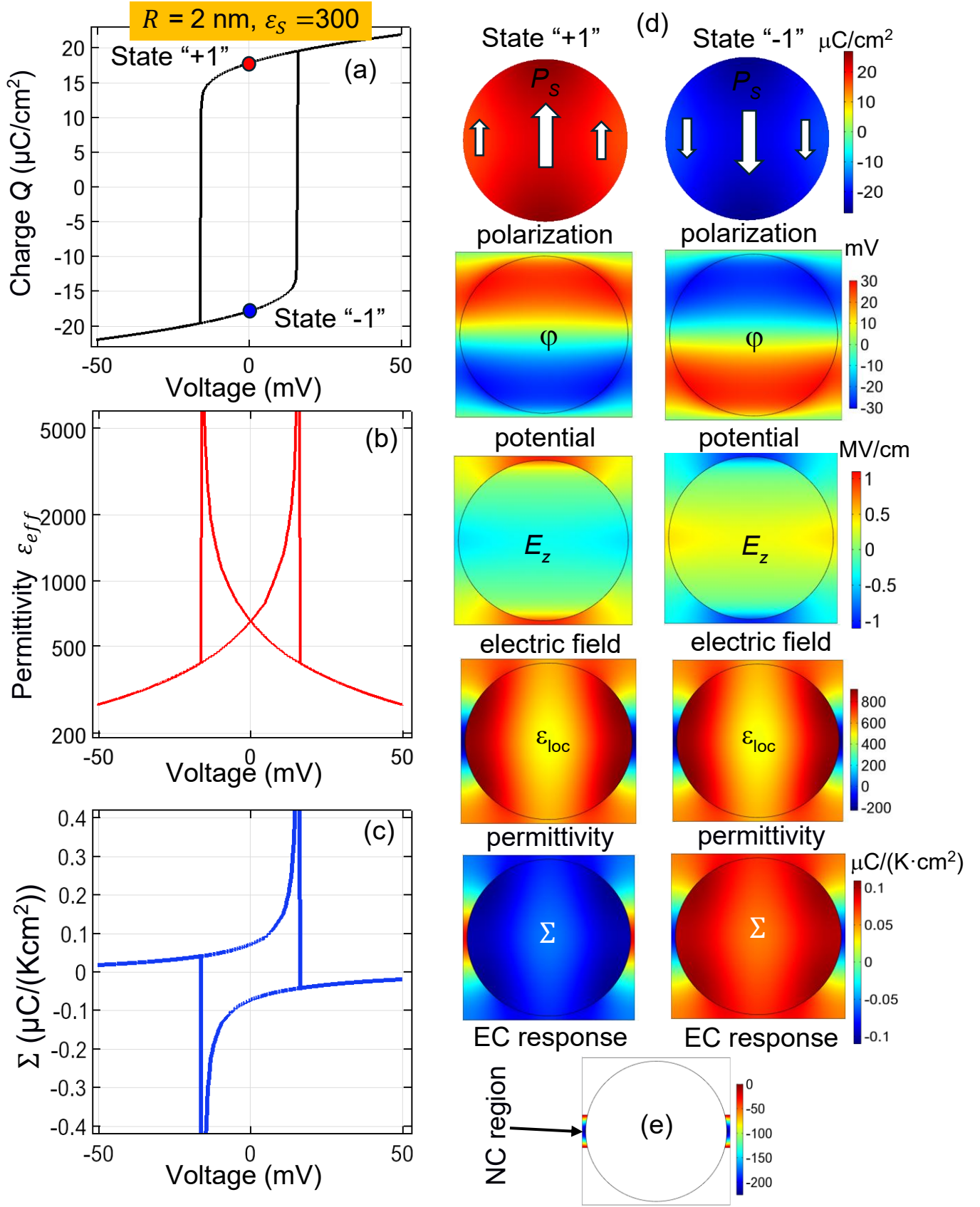

**FIGURE 4.** Dependences of **(a)** charge density on the electrodes, **(b)** effective dielectric permittivity $\varepsilon_{eff}$, and **(c)** EC coefficient $\Sigma$ on the applied voltage. **(d)** Spatial distributions of spontaneous polarization $P_S$, electric potential $\varphi$, internal electric field component $E_z$, local dielectric permittivity $\varepsilon_{loc}(\vec{r})$, and EC response of the BaTiO₃ nanowire in the "+1" (right column) and "-1" (left column) states, as indicated on the charge loop. **(e)**

Spatial regions where $\varepsilon_{loc}(\vec{r})$ is negative. "NC" stands for the negative capacitance. The wire radius $R = 2$ nm, the voltage period is $10^4$ relaxation times, $\varepsilon_S = 300$, and $T = 298$ K.

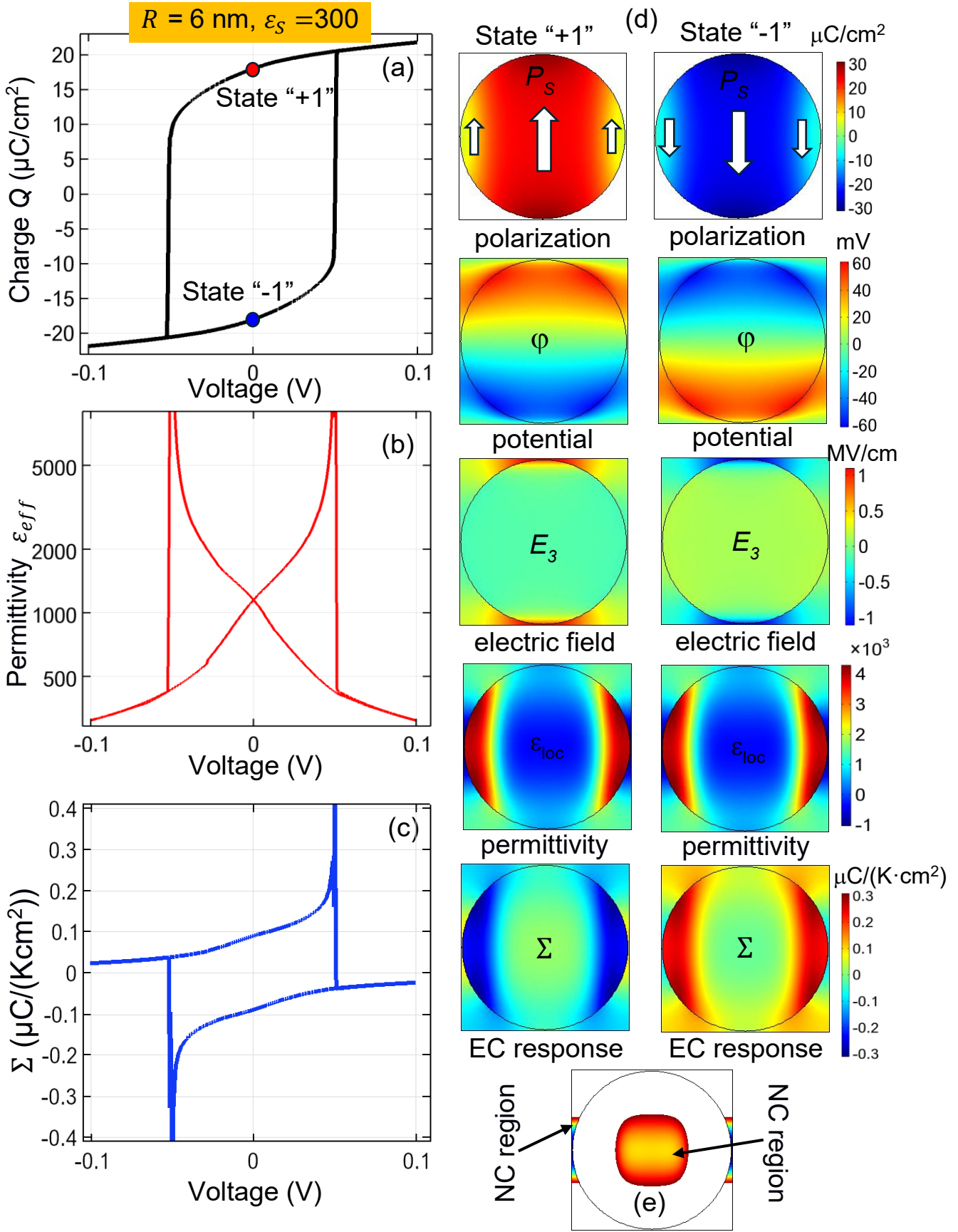

**FIGURE 5.** Dependences of **(a)** charge density on the electrodes, **(b)** effective dielectric permittivity $\varepsilon_{eff}$, and **(c)** EC coefficient $\Sigma$ on the applied voltage. **(d)** Spatial distributions of spontaneous polarization $P_S$, electric potential $\varphi$, internal electric field component $E_z$, local dielectric permittivity $\varepsilon_{loc}(\vec{r})$, and EC response of the BaTiO₃ nanowire in the "+1" (right column) and "-1" (left column) states, as indicated on the charge loop. **(e)** Spatial regions where $\varepsilon_{loc}(\vec{r})$ is negative. "NC" stands for the negative capacitance. The wire radius $R = 6$ nm, the voltage period is $10^4$ relaxation times, $\varepsilon_S = 300$, and $T = 298$ K.

In the next section, we consider potential applications of NC states for the realization of steep slope ferroelectric field-effect transistors (FeFETs).

## IV. NEGATIVE CAPACITANCE STATE IN FERROELECTRIC NANOWIRES ARRAY FOR FeFET APPLICATIONS

Consider a FeFET system in which a layer of densely packed ferroelectric nanowires is sandwiched between the upper gate electrode and a 2D conductive channel with a small effective screening length $\lambda$ (e.g., a channel made from a graphene or a transition metal dichalcogenide). The space between the nanowires is filled with an elastically soft dielectric with a high permittivity $\varepsilon_s \approx \varepsilon_f$. The 2D channel is separated from the bottom gate electrode by a dielectric oxide layer of thickness $d$ and a relative dielectric permittivity $\varepsilon_d$ (**Fig. 6(a)**). We assume that a ferroelectric nanowire is a wide-bandgap material close to a dielectric, and therefore the upper gate is electrically isolated from the channel. The key difference between the geometry shown in **Fig. 6(a)** and the layered geometry considered in Ref. [36] lies in the following: the thickness of the ferroelectric layer is rigidly defined by the nanowire diameter, $2R$. We consider a very long channel of length $L$ containing many nanowires ($L \gg 2R$). It is important to note that this system cannot be treated as a series connection of elements with positive and negative permittivity (as considered in Ref. [36]), because the wire cores with a negative dielectric permittivity are surrounded by thick shells with positive dielectric permittivity.

The subthreshold swing $S$ is a fundamental characteristic of FETs; the device geometry and gate-terminal configuration are illustrated in **Fig. 6(a)**. $S$ quantifies how much the gate voltage $U_g$ must be increased in the subthreshold region to produce a tenfold increase in the drain current $I_d$:

$$S \equiv \ln(10) \frac{dU_g}{d(\ln I_d)}. \tag{1}$$

It is known that in a high-quality transistor with a large gate capacitance, the value of $S$ is limited by thermodynamics and is equal to $S_0 \equiv \ln(10) \frac{k_B T}{e}$, where $e$ is the electron charge, $k_B$ is the Boltzmann constant, and $T$ is the temperature in Kelvin. The importance of the subthreshold swing is that its smallest limiting value $S_0$, the so-called Boltzmann limit, determines the minimum possible operating voltage of the transistor; $S_0 \approx 60$ mV/decade at room temperature. Therefore, reducing $S$ below the fundamental limit $S_0$ offers significant opportunities to reduce power consumption and enable transistor scaling [37, 38].

The algorithm used in the previous work [36] can be adapted to the system shown in **Fig. 6(a)**. However, the inhomogeneous distribution of $\varepsilon_{loc}(\vec{r})$ makes such a simulation cumbersome and precludes analytical solutions, thereby reducing the clarity of the results. Therefore, we replace the array of ferroelectric nanowires, shown in **Fig. 6(a)**, with the simplified model in **Fig. 6(b)**. Within the model, each nanowire consists of a central region with negative local permittivity $\varepsilon_{NC}$ (i.e., the average value of $\varepsilon_{loc}(\vec{r})$ within the NC region), while its outer region with positive local dielectric permittivity $\varepsilon_{loc}(\vec{r})$ is replaced by an effective positive value $\varepsilon_{PC}$. Thus, we use a core-shell model for the wire: the core has a constant negative permittivity $\varepsilon_{NC}$ and its shell has a constant positive permittivity $\varepsilon_{PC}$. To simplify further calculations, we assume that the dielectric environment has approximately the same permittivity as the nanowire shell, i.e., $\varepsilon_s \approx \varepsilon_{PC}$. Using the core-shell model, the effective dielectric permittivity of the system shown in **Fig. 6(b)** can be derived analytically using the effective medium approximation (EMA) (see Refs. [39, 40, 41] and references therein). As shown in **Fig. 6(a)**, this is a rough approximation, but it allows us to use the EMA for a binary mixture.

The EMA allows one to derive an algebraic equation for the effective permittivity $\varepsilon_{eff}$ (see Ref. [42] and **Appendix C** [26] for details). For the system comprised of the cylindrical nanowires polarized perpendicular to their axis, the solution of the EMA equation has the form:

$$\varepsilon_{eff}^{\pm} = \tfrac{1}{2}\big[(\varepsilon_s - \varepsilon_{NC})(1 - 2\mu) \pm \sqrt{(\varepsilon_s - \varepsilon_{NC})^2(1 - 2\mu)^2 + 4\varepsilon_s\varepsilon_{NC}}\,\big], \tag{2}$$

where $\mu$ is the volume fraction ($0 < \mu < 1$) of the regions with $\varepsilon_{NC} < 0$. Since $\varepsilon_{NC}$ is negative and $\varepsilon_s$ is positive, both roots, $\varepsilon_{eff}^-$ and $\varepsilon_{eff}^+$, are positive for $\mu < \tfrac{1}{2}$ and negative for $\mu > \tfrac{1}{2}$.

Under the above assumptions, the nanowire array is approximated by an equivalent layered structure with "effective" dielectric properties, following the Maxwell–Wagner approach that underlies the EMA. This substitution does not imply a true series connection of elements but provides an approximate analytical framework consistent with Eq. (2).

**Analysis of the positive capacitance (PC) states.** Assuming that $\mu \ll \tfrac{1}{2}$, which is indeed the case for $R = 6$ nm and $\varepsilon_s = 300$ (see **Fig. 6(b)**, where $\mu \approx \tfrac{1}{9}$), we derive approximate expressions $\varepsilon_{eff}^+ \approx (\varepsilon_s - \varepsilon_{NC})(1 - 2\mu) + \frac{\varepsilon_s\varepsilon_{NC}}{(\varepsilon_s - \varepsilon_{NC})(1-2\mu)}$ and $\varepsilon_{eff}^- \approx -\frac{\varepsilon_s\varepsilon_{NC}}{(\varepsilon_s - \varepsilon_{NC})(1-2\mu)}$, both of which are positive. The larger root $\varepsilon_{eff}^+$ corresponds to the smaller value of the positive electrostatic energy (proportional to $\frac{1}{\varepsilon_{eff}^{\pm}}$) and therefore corresponds to the thermodynamically stable PC states. For $0 < \mu < \tfrac{1}{2}$, the approximate expression $\frac{1}{\varepsilon_{eff}^-} \approx (1 - 2\mu)\left[\frac{1}{\varepsilon_s} - \frac{1}{\varepsilon_{NC}}\right]$ is smaller than the value $\frac{1}{\varepsilon_s} - \frac{1}{\varepsilon_{NC}}$, which corresponds to two capacitors of equal thickness with permittivities $\varepsilon_s$ and $-\varepsilon_{NC}$ connected in series. Since we consider $\varepsilon_{eff}^+ > \varepsilon_{eff}^-$, we conclude that both roots, $\varepsilon_{eff}^+$ and $\varepsilon_{eff}^-$, are greater than the permittivity of two dielectric regions (with permittivity $\varepsilon_s$ and $\varepsilon_{NC}$) connected in series. As anticipated, the PC states allow the subthreshold swing to approach the Boltzmann limit, although they do not permit it to fall below the limiting value $S_0$.

**Analysis of the negative capacitance (NC) states.** Assuming that the case $\frac{1}{2} < \mu < 1$ can be realized, the NC state appears in large areas of the ferroelectric nanowires, where $\varepsilon_{NC} < 0$. In this regime, the effective capacitance of the nanowire array between the top gate and the conducting channel becomes negative, with $|\varepsilon_{eff}^-| > |\varepsilon_{eff}^+|$ for $\frac{1}{2} < \mu < 1$. The subthreshold swing is reduced according to the expressions derived in Ref. [36]:

$$\frac{S}{S_0} = \left(1 + \frac{C_d}{C_f}\right) \approx 1 + \frac{\frac{1}{2\lambda} + \frac{\varepsilon_d}{d}}{\frac{1}{2\lambda} + \frac{\varepsilon_{eff}}{2R}}. \tag{3}$$

Here, $C_f$ is the differential capacitance of the ferroelectric layer and the channel calculated for zero drain current, and $C_d$ is the differential capacitance of the dielectric oxide (see **Fig. 6(c)**). Thus, the inequality $\varepsilon_{eff}^- < -\frac{R}{\lambda}$ defines the condition for reducing the subthreshold swing below the Boltzmann limit, which can be of great importance for the scaling of FeFETs.

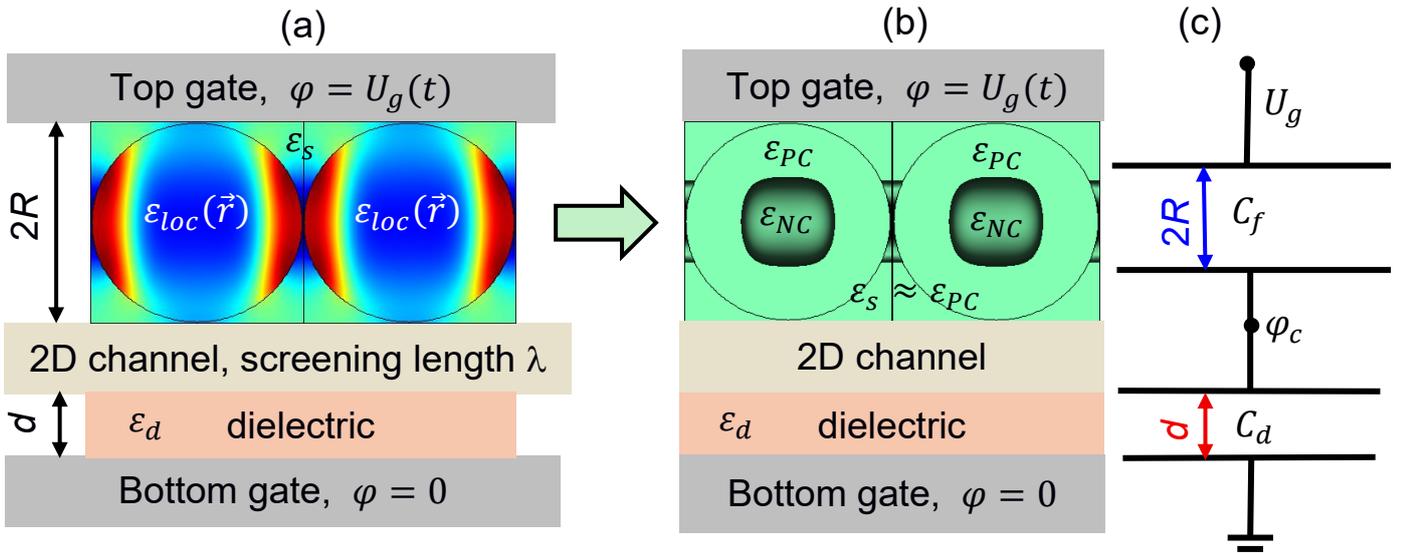

**FIGURE 6. (a)** Transistor structure: top gate, ferroelectric nanowires in the dielectric matrix, 2D conducting channel, dielectric, and bottom gate. **(b)** Simplified model of structure (a) described in the text. **(c)** The equivalent scheme of the NC FeFET.

## V. CONCLUSIONS

We considered a one-dimensional array of uniformly oriented ferroelectric nanowires with circular cross-sections, densely packed between flat electrodes. The spontaneous polarization of each wire is normal to its axis of radial symmetry. The electric potential, electric field, spontaneous polarization, emerging domain structures, dielectric permittivity, and EC response of the nanowires were calculated using the LGD approach with electrostatic equations and elasticity theory.

Due to the size and depolarization effects, the paraelectric, poly-domain, or single-domain ferroelectric states of the nanowires can be stable depending on their radius $R$ and the environment dielectric permittivity $\varepsilon_S$. Both parameters, $R$ and $\varepsilon_S$, determine the phase state of the nanowire through their influence on the depolarization field produced by the nanowire sandwiched between the electrodes. This conclusion qualitatively agrees with that of Ref. [2] for spherical BaTiO₃ nanoparticles in an isotropic dielectric liquid (heptane), where it was shown that the size and concentration of the nanoparticles determine the suspension's phase state. Note that the relative contribution of the dipole-dipole interactions and the effective dielectric permittivity of the suspension increase with an increase in the particle concentration, which makes the qualitative agreement with this work clearer.

We underline the key differences between the phase diagrams of a nanowire array and that of an isolated nanowire embedded in the flat capacitor with the same $\varepsilon_S$. Due to the dipole-dipole interactions between the nanowires, the antipolar state of their array is stable for wire sizes significantly smaller than the critical size of the paraelectric transition in the isolated wire. Another important difference is the appearance and stability of a large mixed state region in the nanowire array. In this case, single-domain and poly-domain ferroelectric states coexist in each wire; the average polarization of the whole array is zero. The mixed state is absent for an isolated wire. The stability of the mixed state in the nanowire array corresponds to the system's minimum electrostatic energy.

The size-induced transition to the paraelectric phase is absent for $\varepsilon_S > \varepsilon_f$, where $\varepsilon_f$ is the dielectric permittivity of the bulk ferroelectric material corresponding to the nanowires. The step-like features that appear in the $R$-dependences of polarization, effective dielectric permittivity, and EC response for $\varepsilon_S > \varepsilon_f$ are associated with sudden changes in the domain configuration (e.g., with the emergence of domains and/or changes in their number within the nanowires). Thus, variations in $R$ and/or $\varepsilon_S$ trigger domain restructuring, which causes abrupt changes in the system's free energy. In addition to the step-like features observed for $\varepsilon_S > \varepsilon_f$, the $R$-dependences of permittivity and the EC coefficient exhibit a broad diffuse maximum associated with the gradual transition from single-domain to poly-domain states as the wire radius increases.

The effective dielectric permittivity of the nanowire array, $\varepsilon_{eff}$, can exceed the ambient permittivity $\varepsilon_S$ by several times. This enhancement cannot be explained by the simple mixture law; rather, it is induced by size and depolarization effects in the nanowires, which increase the local dielectric permittivity $\varepsilon_{loc}$ to values much greater than $\varepsilon_f$.

Hysteresis loops of EC response reveal a significant enhancement of the EC coefficient $\Sigma$, both near the coercive voltages and at small voltages. Near the coercive voltages, $\Sigma$ exceeds 0.3 μC/(Kcm²) for $R = 2$ nm and $\varepsilon_S = 300$. At small voltages $U \approx 0$, the value of $\Sigma$ reaches 0.1 μC/(Kcm²) for $R = 6$ nm and $\varepsilon_S = 300$. This value far exceeds the value of $\Sigma$ for bulk BaTiO$_3$, which is about 0.02 μC/(Kcm$^2$) for small voltages. The strong enhancement of the EC response is driven by the combined action of size effects in single-domain nanowires and electrostatic interactions between the bound charges and image charges in the electrodes.

The spatial distribution of the local dielectric permittivity $\varepsilon_{loc}(\vec{r})$ inside the wires is determined by the distribution of spontaneous polarization: it is maximal in the spatial regions, where the polarization changes most strongly, and minimal in the spatial regions, where the variation is weakest. Thus, when the NC state with $\varepsilon_{loc}(\vec{r}) < 0$ occurs in certain areas of ferroelectric nanowires at $\varepsilon_S \geq \varepsilon_f$, the effective capacitance of the nanowire array between the top gate and the conducting channel increases for $0 < \mu < \frac{1}{2}$ or becomes negative for $\frac{1}{2} < \mu < 1$, where $\mu$ is the volume fraction of the NC regions. According to the conclusions of Ref. [36], the dominance of the NC states creates conditions for reducing the subthreshold swing below the Boltzmann limit, which may be of great importance for the scaling of FeFETs and the reduction of power consumption.

The appropriate selection of the dielectric permittivity of the environment and the radius of the nanowires makes it possible to achieve the maximal NC effect. It is also possible to achieve the maximal enhancement of the EC response due to the size effects in the nanowires. The underlying physics of the predicted enhancement is the combined action of size effects and long-range electrostatic interactions between the ferroelectric dipoles in the nanowires and the image charges in the electrodes.

**Acknowledgements.** The work of A.N.M. is funded by the EOARD project 9IOE063 (related STCU partner project is P751b) and sponsored by the NATO Science for Peace and Security Program under grant SPS G5980 "FRAPCOM". The work of E.A.E. is funded by the National Research Foundation of Ukraine (project "Silicon-compatible ferroelectric nanocomposites for electronics and sensors," grant N 2023.03/0127). The work of O.V.B. is funded by the NAS of Ukraine, grant No. 07/01-2025(6) "Nano-sized multiferroics with improved magnetocaloric properties." Results were visualized in Mathematica 14.0 [43]. The work of O.S.P. is funded by the Ministry of Science and Education of Ukraine, contract M/35-2025 "Flexible nano-ferroelectrics for rapid cooling of combat electronics."

**Authors' contribution.** A.N.M. and M.V.S. generated the research idea and wrote the manuscript draft. A.N.M. formulated the problem and performed analytical calculations. O.V.B. and E.A.E. wrote the codes and performed numerical modelling. O.S.P. performed data processing. D.R.E. worked on the results analysis, explanation, and manuscript improvement. Z.K. and D.R.E. coordinated the research.

# Supporting Information

## APPENDIX A. The LGD Free Energy Functional

The geometry of the nanowire array is shown in **Fig. 1(a)** of the main text. The electric displacement vector has the form $\boldsymbol{D} = \varepsilon_0 \boldsymbol{E} + \boldsymbol{P}$ inside the nanowires. In this expression, $\boldsymbol{P}$ is an electric polarization containing the spontaneous and field-induced contributions. The expression $D_i = \varepsilon_0 \varepsilon_S E_i$ is valid in the dielectric surrounding.

The electric field components, $E_i$, are derived from the electric potential $\varphi$ in a conventional way, $E_i = -\partial \varphi / \partial x_i$. The potential $\varphi_f$ satisfies the Poisson equation inside each of the ferroelectric nanowires (subscript "$f$"):

$$\varepsilon_0 \varepsilon_b \left( \frac{\partial^2}{\partial x_1^2} + \frac{\partial^2}{\partial x_2^2} + \frac{\partial^2}{\partial x_3^2} \right) \varphi_f = \frac{\partial P_i}{\partial x_i}. \tag{A.1a}$$

Here, $\varepsilon_0$ is the vacuum permittivity and $\varepsilon_b$ is a relative background permittivity of the ferroelectric [44]. As a rule, $4 < \varepsilon_b < 10$.

The electric potential $\varphi_S$ in the dielectric surrounding outside the wires satisfies the Laplace equation (subscript "$s$"):

$$\varepsilon_0 \varepsilon_S \left( \frac{\partial^2}{\partial x_1^2} + \frac{\partial^2}{\partial x_2^2} + \frac{\partial^2}{\partial x_3^2} \right) \varphi_S = 0. \tag{A.1b}$$

Equations (A.1) are supplemented with the continuity conditions for the electric potential and normal components of the electric displacements at the wire surface $S_W$:

$$\left. \left( \varphi_S - \varphi_f \right) \right|_{S_W} = 0, \quad \left. \vec{n} (\vec{D}_S - \vec{D}_f) \right|_{S_W} = 0. \tag{A.1c}$$

Electric charges are absent, and the applied voltage is fixed at the boundaries of the computation region:

$$\varphi_S|_{z=0} = 0, \quad \varphi_S|_{z=2R} = U(t). \tag{A.1d}$$

Periodic boundary conditions for electric displacement and potential are applied in the transverse direction to model the wire array.

The LGD free energy functional $G$ of the nanowire polarization $\boldsymbol{P}$ includes a Landau expansion of the 2-nd, 4-th, 6-th, and 8-th powers of the polarization ($G_{Landau}$), a polarization gradient energy contribution ($G_{grad}$), an electrostatic contribution ($G_{el}$), the elastic, linear, and nonlinear electrostriction couplings and flexoelectric contributions ($G_{es+flexo}$), and a surface energy ($G_S$). The free energy functional $G$ has the form [45, 46, 47]:

$$G = G_{Landau} + G_{grad} + G_{el} + G_{es+flexo} + G_{VS} + G_S, \tag{A.2}$$

$$G_{Landau} = \int_{0 < r < R} d^3 r \left[ a_i P_i^2 + a_{ij} P_i^2 P_j^2 + a_{ijk} P_i^2 P_j^2 P_k^2 + a_{ijkl} P_i^2 P_j^2 P_k^2 P_l^2 \right], \tag{A.3a}$$

$$G_{grad} = \int_{0 < r < R} d^3 r \frac{g_{ijkl}}{2} \frac{\partial P_i}{\partial x_j} \frac{\partial P_k}{\partial x_l}, \tag{A.3b}$$

$$G_{el} = - \int_{0 < r < R} d^3 r \left( P_i E_i + \frac{\varepsilon_0 \varepsilon_b}{2} E_i E_i \right) - \frac{\varepsilon_0}{2} \int_{r > R} \varepsilon_{ij}^S E_i E_j d^3 r, \tag{A.3c}$$

$$G_{es+flexo} = -\int_{0<r<R} d^3r \left( \frac{s_{ijkl}}{2} \sigma_{ij}\sigma_{kl} + Q_{ijkl}\sigma_{ij}P_kP_l + Z_{ijklmn}\sigma_{ij}P_kP_lP_mP_n + \right.$$

$$\left. \frac{1}{2} W_{ijklmn}\sigma_{ij}\sigma_{kl}P_mP_n + F_{ijkl}\sigma_{ij}\frac{\partial P_l}{\partial x_k} \right), \tag{A.3d}$$

$$G_S = \frac{1}{2}\int_{r=R} d^2r \, a_{ij}^{(S)} P_iP_j. \tag{A.3e}$$

The coefficient $a_i$ linearly depends on temperature $T$,

$$a_i(T) = \alpha_T[T - T_C], \tag{A.4}$$

where $\alpha_T$ is the inverse Curie-Weiss constant and $T_C$ is the ferroelectric Curie temperature. LGD coefficients and other material parameters of a BaTiO$_3$ are listed in **Table A1**.

According to Maxwell relations, the electrocaloric (EC) coefficient, defined as the isothermal entropy change with electric field $E$, is equal to the pyroelectric effect, i.e., $\Sigma = \left(\frac{\partial \Delta S}{\partial E}\right)_T = \left(\frac{\partial P}{\partial T}\right)_E$. Using the definition of the pyroelectric coefficient, $\Pi(T,E) = -\left(\frac{\partial P}{\partial T}\right)_E$, it is possible to estimate the EC temperature change $\Delta T_{EC}$ and EC response $\Sigma(E)$ as:

$$\Delta T_{EC}(E) \cong -T\int_0^E \frac{1}{\rho_P C_P}\Pi(T,E) \, dE, \tag{A.5a}$$

$$\Sigma(E) \cong -\Pi(T,E) = \left(\frac{\partial P}{\partial T}\right)_E. \tag{A.5b}$$

Here, $\rho_P$ is the mass density, $C_P$ is the heat capacity at constant pressure, and the field-dependent polarization $P(T,E)$ is substituted for the spontaneous polarization $P_S$.

**Table AI.** LGD coefficients and other material parameters of a BaTiO$_3$ in Voigt notations. Adapted from Ref.[48].

| Parameter, its description, and dimension | The numerical value or variation range of the LGD parameters |
|---|---|
| Expansion coefficients $a_i$ in the term $a_iP_i^2$ in Eq.(A.2b) (C$^{-2}$·mJ) | $a_1 = 3.33(T-383)\times10^5$ |
| Expansion coefficients $a_{ij}$ in the term $a_{ij}P_i^2P_j^2$ in Eq.(A.2b) (C$^{-4}$·m$^5$J) | $a_{11} = 3.6\ (T-448)\times10^6$, $a_{12} = 4.9\times10^8$ |
| Expansion coefficients $a_{ijk}$ in the term $a_{ijk}P_i^2P_j^2P_k^2$ in Eq.(A.2b) (C$^{-6}$·m$^9$J) | $a_{111} = 6.6\times10^9$, $a_{112} = 2.9\times10^9$, $a_{123} = 3.64\times10^{10}+7.6(T-293)\times10^{10}$. |
| Expansion coefficients $a_{ijkl}$ in the term $a_{ijkl}P_i^2P_j^2P_k^2P_l^2$ in Eq.(A.2b) (C$^{-8}$·m$^{13}$J) | $a_{1111} = 4.84\times10^7$, $a_{1112} = 2.53\times10^7$, $a_{1122} = 2.80\times10^7$, $a_{123} = 9.35\times10^7$. |
| Linear electrostriction tensor $Q_{ijkl}$ in the term $Q_{ijkl}\sigma_{ij}P_kP_l$ in Eq.(A.2e) (C$^{-2}$·m$^4$) | In Voigt notations $Q_{ijkl} \rightarrow Q_{ij}$, which are equal to $Q_{11}$=0.11, $Q_{12}= -0.045$, $Q_{44}$=0.059 |
| Nonlinear electrostriction tensor $Z_{ijklmn}$ in the term $Z_{ijklmn}\sigma_{ij}P_kP_lP_mP_n$ in Eq.(A.2e) (C$^{-4}$·m$^8$) | In Voigt notations $Z_{ijklmn} \rightarrow Z_{ijk}$. $Z_{ijk}$ varies in the range $-1 \leq Z_c \leq 1$ [49] |
| Nonlinear electrostriction tensor $W_{ijklmn}$ in the term $W_{ijklmn}\ \sigma_{ij}\sigma_{kl}P_mP_n$ in Eq.(A.2e) (C$^{-2}$·m$^4$ Pa$^{-1}$) | In Voigt notations $W_{ijklmn} \rightarrow W_{ijk}$. $W_{ij3}$ varies in the range of $0 \leq W_c \leq 10^{-12}$ as a very small free parameter, and we can neglect it, $W_{ij3} = 0$ |
| Elastic compliance tensor, $s_{ijkl}$, in Eq.(A.2e) (Pa$^{-1}$) | In Voigt notations $s_{ijkl} \rightarrow s_{ij}$, which are equal to $s_{11}$=8.3×10$^{-12}$, $s_{12}$=−2.7×10$^{-12}$, $s_{44}$=9.24×10$^{-12}$. |
| Polarization gradient coefficients $g_{ijkl}$ in Eq.(A.2c) (C$^{-2}$m$^3$J) | In Voigt notations $g_{ijkl} \rightarrow g_{ij}$, which are equal to: g$_{11}$=1.0×10$^{-10}$, g$_{12}$= 0.3×10$^{-10}$, g$_{44}$= 0.2×10$^{-10}$. |
| Surface energy coefficients $a_{ij}^{(S)}$ in Eq.(A.2f) | 0 (corresponding to the natural boundary conditions) |
| Core radius $R_c$ (nm) | Variable: 5 − 50 |
| Background permittivity $\varepsilon_b$ in Eq.(A.2d) (unity) | 7 |

* $\alpha = 2a_1$, $\beta = 4a_{11}$, $\gamma = 6a_{111}$, and $\delta = 8a_{1111}$

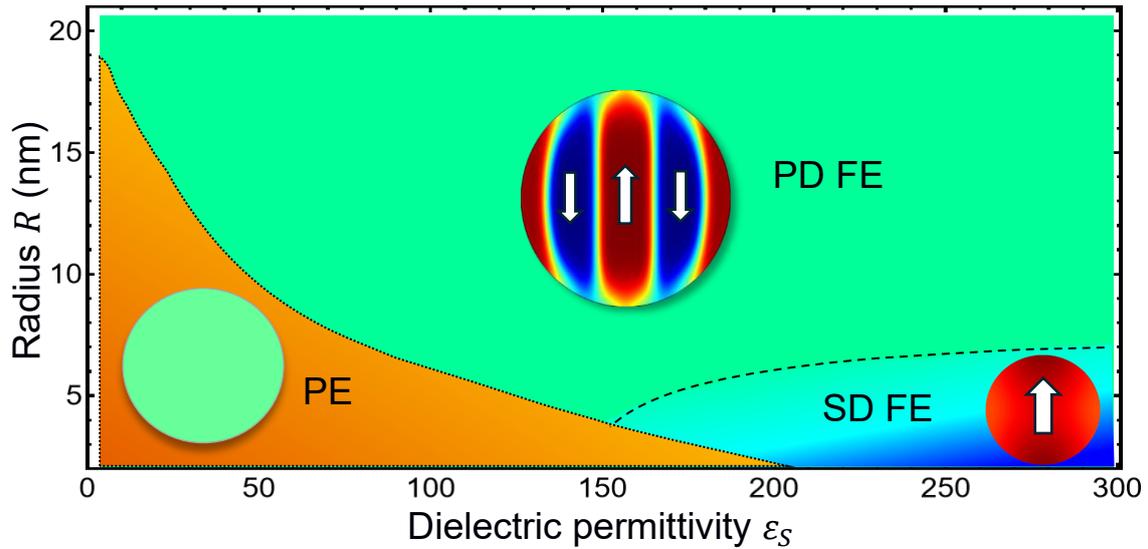

**FIGURE S1.** The ground state of an isolated BaTiO$_3$ nanowire as a function of wire radius $R$ and relative dielectric permittivity of the environment $\varepsilon_S$ for $U = 0$ and room temperature $T = 298$ K. The abbreviations PE, PD FE, and SD FE denote the paraelectric, polydomain ferroelectric, and single-domain ferroelectric states, respectively. Color images with white arrows schematically illustrate the spontaneous polarization distribution and direction in the XZ cross-section of the nanowire for paraelectric, polydomain, and single-domain (up-polarized) states. The color scale for polarization is the same as in **Fig. 1(b)**: positive values are red, negative values are blue, and zero values are light green.

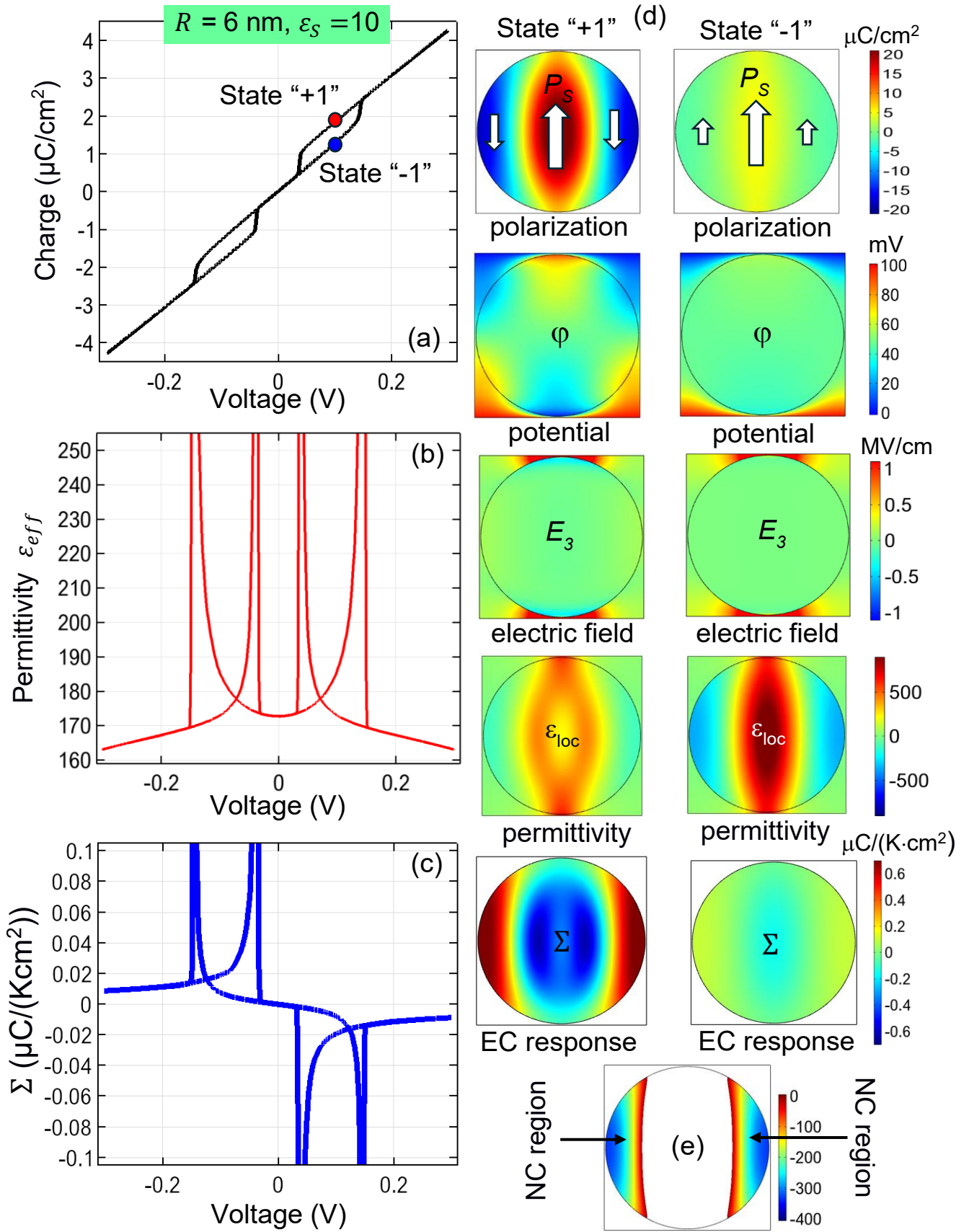

**FIGURE S2.** Dependence of **(a)** the charge density on the electrodes, between which the array of BaTiO₃ nanowires is placed; **(b)** the dielectric permittivity; and **(c)** the EC coefficient Σ on the applied voltage. The panels in part **(d)** from top to bottom are distributions of the spontaneous electric polarization $P_S$; electric

potential $\varphi$; internal electric field; local dielectric permittivity; and EC response of the BaTiO₃ nanowire in states "+1" (right column) and "-1" (left column), which are indicated on the charge loop. Inset **(e)** shows the spatial regions where the local dielectric susceptibility is negative. The wire radius is $R = 6$ nm, the voltage period is $10^4$ relaxation times, $\varepsilon_S = 10$, and $T = 298$ K.

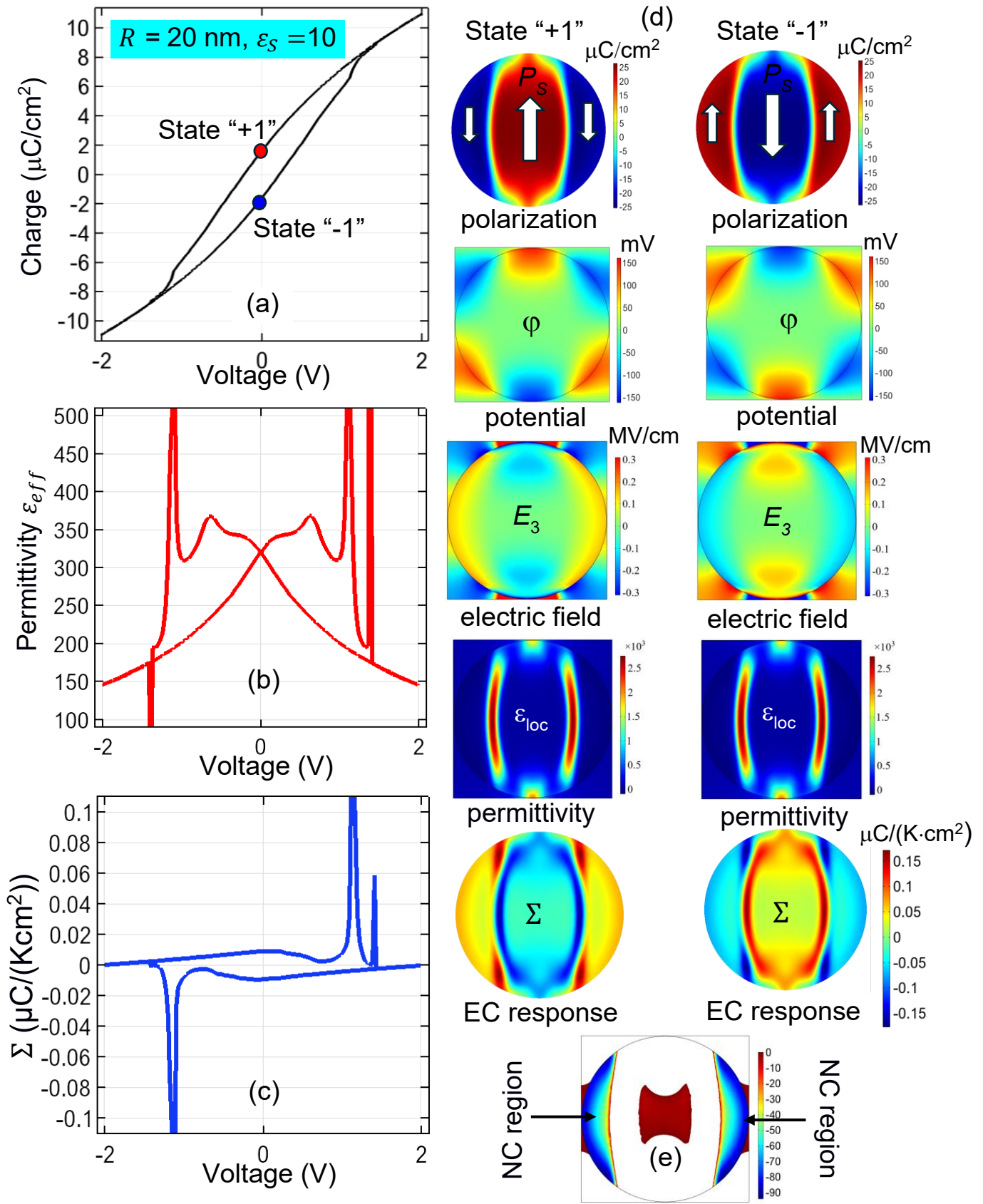

**FIGURE S3.** Dependence of **(a)** the charge density on the electrodes, between which the array of BaTiO$_3$ nanowires is placed; **(b)** the effective dielectric permittivity $\varepsilon_{eff}$; and the EC coefficient $\Sigma$ **(c)** on the applied voltage. The panels in part **(d)** from top to bottom are distributions of the spontaneous electric polarization $P_S$; electric potential $\varphi$; internal electric field; local dielectric permittivity; and EC response of the BaTiO$_3$ nanowire in states "+1" (right column) and "-1" (left column), which are indicated on the charge loop. Inset **(e)** shows the spatial regions where the local dielectric susceptibility is negative. The wire radius is $R = 20$ nm, the voltage period is $10^4$ relaxation times, $\varepsilon_S = 10$, and $T = 298$ K.

## APPENDIX C. Analysis of the EMA Equation

EMA considers the algebraic equation for the effective permittivity $\varepsilon_{eff}$ of the binary mixture:

$$(1 - \mu) \frac{\varepsilon_{eff}^* - \varepsilon_b^*}{(1 - n_a)\varepsilon_{eff}^* + n_a \varepsilon_b^*} + \mu \frac{\varepsilon_{eff}^* - \varepsilon_a^*}{(1 - n_a)\varepsilon_{eff}^* + n_a \varepsilon_a^*} = 0. \tag{C.1}$$

Here $\varepsilon_a^*$, $\varepsilon_b^*$ are complex functions of the relative permittivity of the components "$a$" and "$b$", $\mu$ and $1 - \mu$ are relative volume fractions of the components "$a$" and "$b$", respectively, and $n_a$ is the depolarization field factor for the inclusions of the type "$a$".

For $n_a = 1$ (i.e., for the system consisting of layers perpendicular to the external field), Eq.(C.1) is reduced to the expression for the Maxwell layered dielectric model, $\varepsilon_{eff}^* = \left(\frac{1-\mu}{\varepsilon_b^*} + \frac{\mu}{\varepsilon_a^*}\right)^{-1}$. For $n_a = 0$ (i.e., for the system of columns parallel to the external field), Eq.(C.1) yields $\varepsilon_{eff}^* = (1 - \mu)\varepsilon_b^* + \mu\varepsilon_a^*$, which is equivalent to the system with parallel-connected capacitors with a complex permittivity $\varepsilon_b^*$ and $\varepsilon_a^*$.

For $n_a = 1/2$ (i.e., for the system consisting of the cylindrical nanowires), the effective permittivity is

$$\varepsilon_{eff}^\pm = \frac{1}{2}\left[(\varepsilon_b^* - \varepsilon_a^*)(1 - 2\mu) \pm \sqrt{(\varepsilon_b^* - \varepsilon_a^*)^2(1 - 2\mu)^2 + 4\varepsilon_b^*\varepsilon_a^*}\right]. \tag{C.3}$$